\documentclass[onecolumn, draft, 12pt]{IEEEtran}

\usepackage{amsmath}
\usepackage{amssymb}
\usepackage{amsthm}
\usepackage{tikz}
\usepackage{dblfloatfix}
\usetikzlibrary{positioning}
\usepackage{pgfplots}
\usepackage{bm}
\usepackage{cite}
\usepackage{multirow}

\theoremstyle{definition}

\usetikzlibrary{plotmarks}
\pgfplotsset{compat=newest} 
\pgfplotsset{plot coordinates/math parser=false}
\newlength\figurewidth
\title{Efficient Computation and Covariance Analysis of Geometry-Based Stochastic Channel Models}
\author{Paul Ferrand\thanks{P. Ferrand (paul.ferrand@huawei.com) is with the Mathematical and Algorithmic Sciences Lab, Huawei Technologies France.}}

\definecolor{col1}{rgb}{0.0000,0.4470,0.7410}%
\definecolor{col2}{rgb}{0.8500,0.3250,0.0980}%
\definecolor{col3}{rgb}{0.9290,0.6940,0.1250}%
\definecolor{col4}{rgb}{0.4940,0.1840,0.5560}%
\definecolor{col5}{rgb}{0.4660,0.6740,0.1880}%
\definecolor{col6}{rgb}{0.3010,0.7450,0.9330}%

\begin{document}
\maketitle

\begin{abstract}
In this work, we study a family of wireless channel simulation models called geometry-based stochastic channel models (GBSCMs).
Compared to more complex ray-tracing simulation models, GBSCMs do not require an extensive characterization of the propagation environment to provide wireless channel realizations with adequate spatial and temporal statistics.
The trade-off they achieve between the quality of the simulated channels and the computational complexity makes them popular in standardization bodies.
Using the generic formulation of the GBSCMs, we identify a matrix structure that can be used to improve the performance of their implementations.
Furthermore, this matrix structure allows us to analyze the spatial covariance of the channel realizations.
In accordance to wide-sense stationary and uncorrelated scattering hypotheses, this covariance is static in frequency and does not evolve with user movement.
Finally, we show that convergence may be slow when using GBSCMs to evaluate the performance of covariance-based algorithms, and propose a solution to alleviate this issue.
\end{abstract}

\section{Introduction}
\label{sec:intro}
Channel models in wireless communications can be categorized into two main families: \emph{design} models that capture essential behaviors from the propagation medium against which we can build efficient transceivers, and \emph{simulation} models against which we can evaluate the performance of different technological solutions~\cite{Ferrand2016}.
Among this last family, there are inherent trade-offs to be made between how representative of a scenario is the simulated channel, how flexible is the parametrization, and obviously how computationally complex is the simulation.
Most simulation models are derived from the ray launching paradigm~\cite{Clerckx2012}.
An immediate approach is to use actual maps of an environment and a ray-tracing tool to generate a realistic channel~\cite{Jamsa2015}.
While ray-tracers can provide very accurate channel realizations, they are limited in their ability as simulation models.
They lack generality in the sense that they are restricted to a specific environment and not a more abstract typical communication scenario~\cite{Ferrand2016}.
On top of this, they tend to be computationally complex.

Geometry-based stochastic channel models (GBSCMs) provide a way to alleviate these issues by concentrating on the relationship between the communication endpoints and interacting objects in the simulation space---commonly named \emph{scatterers}.
For a given link between communication endpoints, the interactions of the transmitted waves with the virtual environment formed by the scatterers are used to provide channel realizations.
One can adopt a scatterer-centric approach and consider that scatterers are shared by all the links in the network.
This approach has been popularized by successive COST actions and is frequently named the COST model; a recent specification can be found in~\cite{Liu2012}.
By increasing the number of scatterers and measuring their position in space, one can fall back to ray-tracing and obtain realistic channels matching specific environments~\cite{Jarvelainen2016}.

An alternative is to be more user-centric: the channel model concentrates on the transmission endpoints and independently generate a virtual scattering environment for each link.
The virtual environment complexity can be limited to power delay profiles, as e.g. in~\cite{Saleh1987}.
It can also consider the angular profile of the impinging waves~\cite{Steinbauer2001}.
This is the preferred approach of the original 3GPP Spatial Channel Model (SCM) and its extensions, as well as the WINNER model family~\cite{Narandzic2007,Kyoesti2007}.
Most recent simulation models developed for standardization purposes use user-centric GBSCMs~\cite{3GPP_TR38901_v1430}; new instances of these models for different scenarios also use the same paradigm~\cite{Jamsa2015, Samimi2016}. 

In this work, we concentrate on these user-centric GBSCMs.
We first propose a generic description of the models and an expression to derive the multiple inputs, multiple outputs (MIMO) channel coefficients for a given time and frequency.
Similarly to the RIMAX channel estimation algorithm~\cite{Richter2004}, we separate the spatial term and the time-frequency terms of the channel coefficient in a very compact matrix product.
This enables memory--computation trade-offs where we store intermediate computation results to accelerate the simulation.
It also enables single instruction, multiple data (SIMD) vectorization of the coefficient generation, where the compiler can improve the machine code through short vector instructions~\cite{Patterson1998}.
We show that as a byproduct of this matrix formulation, we can also express the spatial covariance of the channels generated using these models.
Since instantaneous channel state information can prove difficult to obtain in modern wireless systems, spatial statistics obtained over a longer period are expected to play a big role in future wireless communication networks~\cite{Adhikary2014, Ferrand2017,8094949}.
In accordance to wide-sense stationary and uncorrelated scattering (WSSUS) principles~\cite{Clerckx2012}, we show that the covariance is static in frequency and does not evolve with user movement.
This provides additional justification to improve user-centric GBSCMs with more covariance consistency, as detailed e.g. in~\cite{Jaeckel2014, Martinez2016a} and in extensions of~\cite{3GPP_TR38901_v1430}.
Finally, we identify that in practical use of such models in simulation, the generated channel might not match the behavior one might expect from its spatial covariance.
We provide an approach to help alleviate this problem.

\begin{figure*}[b!]
\hrule
\normalsize
\newcounter{MYtempeqncnt}
\setcounter{MYtempeqncnt}{\value{equation}}
\setcounter{equation}{2}
\begin{equation}
\label{eq:coeff}
\begin{split}
	h_{r,s}(f, t) = \sum_{n = 1}^N \sum_{m = 1}^{M_n} &\sqrt{P_{n, m}} F_s\left(\theta^{(D)}_{n,m}, \phi^{(D)}_{n,m} \right) \exp [ \jmath \psi_{n,m} ] F_r\left(\theta^{(A)}_{n,m}, \phi^{(A)}_{n,m} \right) \\ &\exp\left[ \jmath k_0  \mathbf{p}^T_s\mathbf d_{n, m} \right] \exp\left[ \jmath k_0 \mathbf{p}^T_r\mathbf a_{n, m}\right] \exp\left[ \jmath k_0 \nu_{n, m} t\right] \exp\left[-\jmath2\pi f\tau_{n, m}\right]
\end{split}
\end{equation}
\setcounter{equation}{\value{MYtempeqncnt}}
\end{figure*}

\section{Channel coefficient generation method}

\begin{figure}[t!]
	\centering

\usetikzlibrary{decorations.pathreplacing}
\begin{tikzpicture}[>=latex, bd/.style={draw, rounded corners, text width=2cm, minimum height =1cm,align=center}]
	\scriptsize
	\node (A) at (0, 0) [bd] {Set the scenario parameters};
	\node (B) [right=1 of A, bd] {Drop the base stations and users};
	\node (C) [right=1 of B, bd] {Assign large scale parameters};
	\node (D) [below=1 of C, bd] {Draw cluster angles, delays and power};
	\node (E) [left=1 of D, bd] {Draw subpath angles, delays and power};
	\node (F) [left=1 of E, bd] {Draw initial phases};
	\draw[->] (A) -- (B);
	\draw[->] (B) -- (C);
	\draw[->] (C) -- (D);
	\draw[->] (D) -- (E);
	\draw[->] (E) -- (F);
	\draw [decorate, decoration={brace,amplitude=10pt, raise=2pt}] (A.north west) -- (C.north east)  node [black,midway,yshift=0.6cm] {Per-link parameters};
	\draw [decorate, decoration={brace,amplitude=10pt, raise=2pt}] (D.south east) -- (D.south west)  node [black,midway,yshift=-0.6cm] {Per-cluster parameters};
	\draw [decorate, decoration={brace,amplitude=10pt, raise=2pt}] (E.south east) -- (F.south west)  node [black,midway,yshift=-0.6cm] {Per-subpath parameters};
\end{tikzpicture}
	\caption{Generic coefficient parameter generation for geometry-based stochastic channel models.}
	\label{fig:gbscm_bg}
\end{figure}
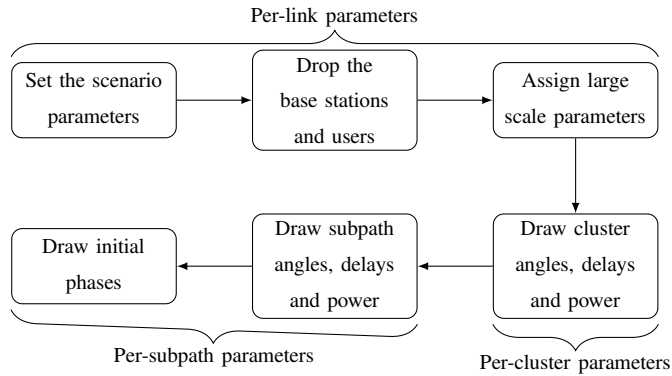

Most channel simulation tools applying the GBSCM paradigm follow a procedure similar to the one described in Fig.~\ref{fig:gbscm_bg} to generate the parameters of the channels between endpoints.
Initial steps aim at setting up the physical simulation environment, as well as the global parameters governing the channel realizations.
The modeler sets the general scenario and the configuration parameters, as well as antenna parameters for all the endpoints in the network.
The channel simulation tool drops the transmitters and the receivers on the 3-dimensional simulation space.
It then assigns large-scale parameters---such as path loss, shadowing, the number of clusters, and second-order statistics about the angular and delay distribution---to the links based on the endpoint positions in space; these parameters may be spatially correlated with those of other links in the network.
After this last step, each link is handled separately by the channel simulation; the dependence between the different links in the network is thus limited to correlation in the large-scale parameters.

For each link, the simulation generates a number of virtual clusters of multipath components.
Each cluster has a mean delay, power, and angles of arrival and departure.
All are drawn at random from specified distributions.
The clusters are then refined into subpaths, with their own power, delay, and angular distributions.
This last step is usually independent between all the clusters and all the links.
For each subpath, initial phases are finally drawn at random.
Note that we do not consider polarization of the antenna arrays to simplify the expression and analysis.
In the appendix, we show that the conclusions of this work can be readily extended to a model including polarized antenna elements.

At the end of the procedure in Fig.~\ref{fig:gbscm_bg}, each link is described by a collection of multipath components grouped into clusters.
Using this description, the channel simulation tool can generate the channel coefficients as follows.
Consider a link between a transmitter and a receiver, and assume without loss of generality the Cartesian and spherical coordinate parameterization of~\cite{3GPP_TR38901_v1430}.
The receiver is assigned a velocity vector $\mathbf v_r$ and the transmitter a velocity $\mathbf v_s$.
Let $1 \leq s \leq S$ and $1 \leq r \leq R$ index the transmitter and receiver antennas in their respective arrays of $S$ and $R$ antenna elements.
The position of the antenna elements in Cartesian coordinates are denoted as $\mathbf{p}_s \in \mathbb R^{3\times 1}$ and $\mathbf{p}_r \in \mathbb R^{3\times 1}$ for the transmit and receive array respectively.
For both the positions and responses of the antenna elements, we let the subscript indicate whether the transmit or receive array is concerned, to lighten the notation.
Each antenna element has a given angular response.
The angular response allocates scalar gains $F_s(\theta, \phi)$ and $F_r(\theta, \phi)$ to impinging rays departing and arriving at an elevation angle $\theta$ and an azimuth angle $\phi$.
The link is described through $N$ clusters, with each cluster comprising $M_n$ subpaths.
Let $1 \leq n \leq N$ index the clusters and $1 \leq m \leq M_n$ the subpaths of the $n^\textrm{th}$ cluster.
Each subpath is characterized by:
\begin{itemize}
	\item a power $P_{n, m}$,
	\item elevation and azimuth angles of arrival $\theta^{(A)}_{n,m}$ and $\phi^{(A)}_{n,m}$,
	\item elevation and azimuth angles of departure $\theta^{(D)}_{n,m}$ and $\phi^{(D)}_{n,m}$,
	\item a delay $\tau_{n,m}$, and
	\item an initial phase $\psi_{n, m}$.
\end{itemize}
For each subpath, we define the departure unit vector as 
\begin{equation}
	\mathbf d_{n, m} = \begin{pmatrix}
		\sin\left(\theta^{(D)}_{n,m}\right) \cos\left(\phi^{(D)}_{n,m}\right)\\
		\sin\left(\theta^{(D)}_{n,m}\right) \sin\left(\phi^{(D)}_{n,m}\right)\\
		\cos\left(\theta^{(D)}_{n,m}\right) 
	\end{pmatrix}.
	\end{equation}
The arrival unit vector $\mathbf a_{n, m}$ is defined similarly using $\theta^{(A)}_{n,m}$ and $\phi^{(A)}_{n,m}$.
We assume a center frequency $f_0$, and define the center wave number as $k_0 = 2\pi f_0/c$, where $c$ is the speed of light.
Finally, we define the Doppler coefficient of each subpath as
\begin{equation}
	\nu_{n, m} = \mathbf a^T_{n, m} \mathbf v_r + \mathbf d^T_{n, m} \mathbf v_s.
\end{equation}
The final frequency response of the link at a given time $t$ and frequency $f$ is denoted by $\mathbf H(f, t) \in \mathbb C^{R \times S}$; each component $H_{r, s}(f, t)$ of this matrix can then be computed as~\eqref{eq:coeff} on the bottom of the page.
\setcounter{equation}{3}

This formulation is quite generic and can be applied to many models.
It subsumes most models following the double-directional paradigm~\cite{Steinbauer2001}, including early models for the 3GPP standardization (SCM and SCME), the WINNER family of models \cite{Kyoesti2007,Narandzic2007}, up to the most recent channel model proposals from standardization bodies such as the GBSCM of METIS~\cite{Jamsa2015} and the latest from the 3GPP~\cite{3GPP_TR38901_v1430}.
It also encompasses more specialized channel models such as the one proposed by NYUSIM for mmWave channels~\cite{Samimi2016}.
It could also be used to approximate the results of more scatterer-centric models such as the COST models~\cite{Liu2012} under some stationarity conditions on the environment.
Depending on the desired application and frequency range, the above references can provide the relevant procedures used to generate the parameters of~\eqref{eq:coeff}.

\section{Matrix formulation}
Our immediate goal is to efficiently generate the channel coefficients over time and frequency for all channel simulation models described by~\eqref{eq:coeff}.
To this aim, we convert \eqref{eq:coeff} into a matrix product and describe how to extract memory--computation trade-offs from it.
First, notice that there is no use in considering individual clusters for coefficient generation: once the multipath components are generated, we can consider them all at the subpath level.
Let $M = \sum_{n=1}^N M_n$ be the total number of subpaths, now indexed solely by $m$.
The channel between transmit antenna $s$ and receive antenna $r$ is generated as a sum over all $M$ subpaths of a product can be factored into 3 main components:
\begin{enumerate}
	\item a factor that depends on the transmit antenna response, transmit antenna position and departure vector,
	\item a factor that depends on the receive antenna response, receive antenna position and arrival vector, and
	\item a factor that depends on the power, phase, Doppler coefficient and delay of the subpath.
\end{enumerate}
We can see that the first component is indexed only by $s$ and $m$, the second one by $r$ and $m$, and the last one only by $m$.
Let us define the spatial transmit matrix $\mathbf S = [S_{s, m}] \in \mathbb C^{S\times M}$ as 
\begin{equation}
	S_{s, m} = F_s\left(\theta^{(D)}_{m}, \phi^{(D)}_{m} \right)\exp\left[ \jmath k_0  \mathbf{p}^T_s\mathbf d_{m} \right]
\end{equation}
and the spatial receive matrix $\mathbf R = [R_{s, m}] \in \mathbb C^{R\times M}$ as
\begin{equation}
	R_{r, m} = F_r\left(\theta^{(A)}_{m}, \phi^{(A)}_{m} \right)\exp\left[ \jmath k_0  \mathbf{p}^T_r\mathbf a_{m} \right].
\end{equation}
Let us then define the phase vector $\bm \psi \in \mathbb C^{M \times 1}$ as
\begin{equation}
	\bm \psi = \left(\exp\left[\jmath\psi_1\right], \ldots, \exp[\jmath\psi_M] \right)^T,
\end{equation}
the power vector $\bm \rho \in \mathbb R^{M \times 1}$ as
\begin{equation}
	\bm \rho =  \left(\sqrt{P_1}, \ldots, \sqrt{P_M}\right)^T,
\end{equation}
the Doppler vector $\bm \nu(t) \in \mathbb C^{M \times 1}$ as
\begin{equation}
	\bm \nu(t) = \left( \exp\left[\jmath k_0 \nu_1 t\right], \ldots, \exp\left[\jmath k_0 \nu_M t\right]\right)^T,
\end{equation}
and the frequency vector $\bm \xi(f) \in \mathbb C^{M \times 1}$ as
\begin{equation}
	\bm \xi(f) = \left( \exp\left[-2\pi\jmath f \tau_1\right], \ldots, \exp\left[-2 \pi \jmath f \tau_M \right]\right)^T.
\end{equation}
The overall channel can be written as follows 
\begin{equation}
	\mathbf H(f, t) = \mathbf R \cdot \mathbf{diag} \, \left( \bm \psi \odot \bm \rho \odot \bm \nu(t) \odot \bm \xi(f) \right) \cdot \mathbf S^T,
\label{eq:matrix_coeffs}
\end{equation}
with $\odot$ denoting the component-wise Hadamard product.

The expression~\eqref{eq:matrix_coeffs} shows that the MIMO channel of each link can be computed in a very efficient way over time and frequency by pre-computing and storing the spatial matrices $\mathbf S$ and $\mathbf R$.
One can weigh each column of $\mathbf S$ or $\mathbf R$ by $\bm \rho$ to further reduce the computation time needed to obtain the channel at a specific time and frequency point.
This formulation also allows compilers and interpreters to efficiently vectorize the computation~\cite{Patterson1998}.
The cost of storing these intermediate results is relatively low: if $L$ is the number of links in the network and assuming all endpoints have the same number of antennas $R$ and $S$, the memory needed to store all the array response matrices is $4 L\times M \times (R + S)$ the size of a real float or double in the architecture.
Note that this formulation may also be derived from the RIMAX formulation of the channel estimation problem~\cite{Richter2004}.
The authors of~\cite{Richter2004} identify a similar spatial matrix structure and combine it with regular sampling in time and frequency to formulate the channel estimation problem as a large non-linear least-squares problem.
As our aim in this paper is to solve the simpler problem of coefficient generation, the expression~\eqref{eq:matrix_coeffs} is more compact and more amenable to computation--memory trade-offs.

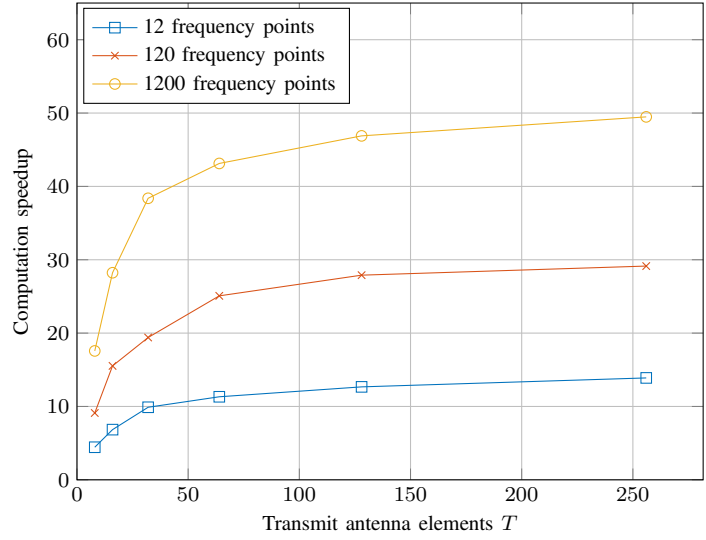
\begin{figure}[t!]
	\scriptsize
	\centering
	\begin{tikzpicture}
	\begin{axis}[
		width=\figurewidth,
		height=0.8\figurewidth,
		xmajorgrids,
		ymajorgrids,
		xlabel={Transmit antenna elements $T$},
		ylabel=Computation speedup,
		legend entries={
			{12 frequency points},
			{120 frequency points},
			{1200 frequency points},
		},
		mark options={solid},
		xmin=0,
		ymin=0,
		ymax=65,
		ytick={0, 10, 20, 30, 40, 50, 60},
		legend style={at={(0.01, 0.99)}, anchor=north west, font=\scriptsize},
		legend cell align=left,
	]
		\addplot+[color=col1, solid, mark=square] coordinates { (8, 4.44) (16, 6.84) (32, 9.89) (64, 11.32) (128, 12.67) (256, 13.89) };
		\addplot+[color=col2, solid, mark=x] coordinates { (8, 9.11) (16, 15.53) (32, 19.40) (64, 25.08) (128, 27.9) (256, 29.14) };
		\addplot+[color=col3, solid, mark=o] coordinates { (8, 17.57) (16, 28.23) (32, 38.38) (64, 43.13) (128, 46.89) (256, 49.48) };
	\end{axis}
	\end{tikzpicture}
	\caption{Computation speedup using the matrix formulation for a single time $t$ over multiple frequencies. Each receiver has $R=4$ antennas, and we vary the number of transmit antennas. Each link has 240 subpaths with parameters generated from~\cite{3GPP_TR38901_v1430} and there are 10 links in the network. }
	\label{fig:computation}
\end{figure}
\begin{table}[b!]
\centering
\caption{Run times in seconds of the baseline coefficient generation~\eqref{eq:coeff} and the optimized one in~\eqref{eq:matrix_coeffs} for a single time $t$ over multiple frequencies for 10 users. The speedup is shown on Fig.\ref{fig:computation}.}
\label{tab:computation}
\begin{tabular}{|c|r|r|r|r|r|r|}
\hline
\# frequency points & \multicolumn{2}{c|}{12} & \multicolumn{2}{c|}{120} & \multicolumn{2}{c|}{1200} \\ \hline
Algorithm& \multicolumn{1}{c|}{Base} & \multicolumn{1}{c|}{Opt.} & \multicolumn{1}{c|}{Base} & \multicolumn{1}{c|}{Opt.} & \multicolumn{1}{c|}{Base} & \multicolumn{1}{c|}{Opt.} \\ \hline
8 antennas  & 0.142 & 0.032 & 1.111 & 0.122 & 10.37 & 0.590 \\ \hline
16 antennas & 0.275 & 0.040 & 2.224 & 0.143 & 22.09 & 0.782 \\ \hline
32 antennas & 0.527 & 0.053 & 4.341 & 0.224 & 42.13 & 1.098 \\ \hline
64 antennas & 0.996 & 0.088 & 8.566 & 0.341 & 78.26 & 1.814 \\ \hline
128 antennas& 2.228 & 0.175 & 17.30 & 0.610 & 156.2 & 3.331 \\ \hline
256 antennas& 4.187 & 0.301 & 35.55 & 1.219 & 317.0 & 6.405 \\ \hline
\end{tabular}
\end{table}
We show the potential gains of SIMD vectorization in Fig.~\ref{fig:computation}.
The simulation model we implemented here is the Urban Micro model from~\cite{3GPP_TR38901_v1430}.
We use a test network scenario with 1 base station and 10 users, and compare the time it takes to obtain the channel for all users over a given number of frequency sub-carriers for both a straightforward implementation of~\eqref{eq:coeff}, and the optimized implementation using~\eqref{eq:matrix_coeffs}.
Both implementation are single-threaded and uses the Eigen C++ library~\cite{eigenweb}, and are compiled on a 2014 laptop with an Intel Core i7 processor.
The nominal run-times of both implementations are transcribed on Tab.~\ref{tab:computation}.
We vary both the number of frequency points polled and the number of antennas at the base station to cover different OFDM frame sizes and different scenarios up to massive MIMO applications.
We compare only the coefficient generation procedure, since the network and parameter set-ups are the same for both computation methods.
Note however that we include the time needed to pre-compute and store the spatial matrices of in the score of the optimized implementation of~\eqref{eq:matrix_coeffs}.
The additional memory needed for the spatial matrices was around 20 megabytes on this processor architecture for the most demanding case with 256 antennas at the transmitter.
The speedup increases as expected with the number of frequency points; for 1200 frequency points, the speedup is almost 18 for only 8 antennas at the transmitter.
This speedup also increases with the number of antennas at the transmitter.


\section{Covariance analysis}

Knowledge of the theoretical transmit and receive covariance of wireless channels is very relevant for modern MIMO physical layer procedures.
It can be used to simplify user scheduling and precoding on large antenna arrays~\cite{Adhikary2014}.
It was also recently shown to be critical in fully avoiding pilot contamination in multi-cell massive MIMO systems, thereby theoretically allowing unbounded capacity~\cite{8094949} under some mild conditions.
When using GBSCMs to validate such algorithms in more practical applications, it is thus necessary to obtain information about the theoretical covariance of the generated channels and study its behavior.

Unlike in scatterer-centric channel models~\cite{Liu2012}, the scatterers are only described through their angle of arrival and departure in user-centric GBSCMs.
These models assume in a way that the impinging waves can be approximated as planar waves.
This simplifies the computation of the channel coefficients because we only need 2 scalar products of arrival and departure vectors with the antenna positions to obtain the array responses.
However, one key issue is that since we do not know the actual position of the scatterers, there is no obvious way to update the arrival and departure vectors when the endpoints are moving---although some approaches have been proposed e.g. in~\cite{Jaeckel2014,Martinez2016a,3GPP_TR38901_v1430}.

The most common user-centric GBSCMs thus tend to assume that for short movements, the arrival and departure vectors are static.
This is clear in \eqref{eq:coeff} and \eqref{eq:matrix_coeffs}, where the only time-dependent component is the Doppler term.
Under the matrix formulation of \eqref{eq:matrix_coeffs}, we can express the transmit and receive covariance from $\mathbf R$, $\mathbf S$ and $\bm \rho$.
We can use this value to evaluate the theoretical performance of algorithms that either rely on covariance information~\cite{Adhikary2014,8094949} or try to estimate this covariance~\cite{Ferrand2017}.
Let us consider the receiver side covariance
\begin{equation}
	\mathbf K_R = \mathbb E_t\left[ \mathbf H^H(f,t) \mathbf H(f,t) \right].
\end{equation}
The expectation here is understood over time.
Let $\mathbf u(f) = \bm \rho \odot \bm \psi \odot \bm \xi(f)$ be the vector of parameters that are independent of time, and let $\mathbf U(f) = \mathbf{diag}\, \mathbf u(f)$ and $\bm V(t) = \mathbf{diag}\, \bm \nu(t)$.
We can expand $\mathbf K_R$ as 
\begin{align}
	\mathbf K_R &= \mathbf R \mathbf U(f) \mathbb E\left[\mathbf V(t) \mathbf S^T \mathbf S^* \mathbf V^H(t)\right]\mathbf U^H(f)  \mathbf R^H.\label{eq:tx_covariance_base}
\end{align}
Recall the following property about complex exponentials
\begin{equation}
	\mathbb E_t\big[\exp[\jmath k_0 t(\nu_m - \nu_{m'})]\big] = \begin{cases}
		0 \quad &\text{if } \nu_m \neq \nu_{m'} \\
		1 \quad &\text{if } \nu_m = \nu_{m'}.
	\end{cases}
	\label{eq:time_average}
\end{equation}
In most practical cases, all the subpaths are going to have distinct Doppler coefficients, since their angles of arrival and angles of departure are going to be different.
We can thus conclude that
\begin{equation}
	\mathbf D = \mathbb E_t\left[\mathbf V(t) \mathbf S^T \mathbf S^* \mathbf V^H(t)\right] \label{eq:inner_covariance_matrix}
\end{equation}
is going to be a diagonal matrix in general.
Now, since we have $\mathbf u(f) \odot \mathbf u^*(f) = \bm \rho \odot \bm \rho$ for all $f$, the transmit covariance of the link can be written as
\begin{equation}
	\mathbf K_R = \mathbf R \cdot \mathbf{diag}\, \bm \rho \cdot \mathbf D \cdot \mathbf{diag}\, \bm \rho \cdot \mathbf R^H. \label{eq:tx_covariance}
\end{equation}
It is indeed independent of $f$, and does not vary upon user movements.
The transmitter side covariance $\bm K_S$ can be computed in a similar way, with the same conclusions.

We can provide some analysis of this expression.
First, we see that such GBSCMs are not suited to evaluate covariance tracking algorithms, which reduce to online covariance estimation since the covariance is not evolving over time.
Second, the convergence of the expectation in~\eqref{eq:time_average} is going to be very slow over time for practical Doppler coefficients.
A rapid check shows that for users moving e.g. at 1 m/s and communicating at 3 GHz, we have that $k_0 |\nu_m| \leq 20 \pi \approx 62.83$.
The sample rate of a typical modern communication system on the other hand is in the order of tens of nanoseconds, and even the length of a complete Long Term Evolution (LTE) frame is in the order of milliseconds~\cite{Clerckx2012}.
We would thus require hundreds of frames to reach the expectation in~\eqref{eq:time_average}.
This may be an issue in practice since the perceived covariance of the channel is going to be very different from the theoretical one computed from~\eqref{eq:tx_covariance}.
For example, the sample covariance computed using successive channel samples or an algorithm such as the one described in~\cite{Ferrand2017} may not be able to converge to the theoretical covariance over a reasonable amount of time.
In a way, the Doppler processes in $\bm \nu(t)$ will only cover a small subspace of the eigenspace of $\mathbf K_R$ over a short amount of time.

One way to alleviate this issue is to randomize the phases in $\bm \psi$ in order to generate an ensemble of channels sharing the same physical environment.
We thus define $\mathbf u'(f) = \bm \rho \odot \bm \xi(f)$ and $\mathbf v'(t) = \bm \psi \odot \bm \nu(t)$, where $\bm \psi$ is now understood as a random vector such that $\mathbb E_{\bm\psi}[\bm \psi \bm \psi^H] = \mathbf I$.
This property is verified in particular when the random phases are uniformly distributed over $[0, 2\pi]$.
All other channel parameters in~\eqref{eq:matrix_coeffs} can and should stay fixed over the ensemble.
The channel is thus not deterministic anymore, and we can take expectations either in time or over the ensemble.
When the expectation is taken over the ensemble, one can verify that $\mathbf D'(t) = \mathbb E_{\bm\psi}\left[\mathbf V'(t) \mathbf S^T \mathbf S^* \mathbf {V'}^{H}(t)\right]$ is going to be a diagonal matrix for all $t$.
Generating channel samples with different random phases can thus accelerate the convergence of the sample covariance to its theoretical value, and in general accelerate the convergence of Monte Carlo simulations.
This solution is similar to the approach taken for general sum-of-sinusoids generators~\cite{Pop2001}.
We stress out however that in this case, the coherence over time of the channel coefficients is destroyed.
Depending on the simulation goals, it may thus be necessary to trade off between the time and ensemble average in order to evaluate the performance of covariance-based algorithms.

\section{Conclusion}
In this work, we provided a compact matrix formulation for the generation of channel matrices used in most user-centric GBSCMs.
This matrix formulation enables performance gains by vectorization and computation--memory trade-offs if the channel is to be generated over multiple time and frequency resources for a given link.
It also allows very straightforward expressions for the spatial covariance of the MIMO channel.
This spatial covariance is shown to be independent of frequency	and does not evolve with user movements.
We finally propose an approach to accelerate the convergence of Monte Carlo simulations for covariance-based algorithms when using channels generated by user-centric GBSCMs.

\appendix
\section{Extension to dual-polarized models}
For polarized antenna elements and channels, the matrix formulation can be found in a similar way.
We consider the following model for a dual-polarized version of a generic GBSCM.
Transmit and receive antenna elements are described by their angular responses $F^V_s(\theta, \phi)$, $F^H_s(\theta, \phi)$, $F^V_r(\theta, \phi)$, $F^V_r(\theta, \phi)$ in the vertical and horizontal polarization domain.
For each subpath, there are now 4 initial phases $\psi_m^{VV}$, $\psi_m^{HH}$, $\psi_m^{HV}$ and $\psi_m^{VH}$, as well as a depolarization coefficient $\kappa_m$.
In \eqref{eq:coeff}, the factors related to the antennas and phases are replaced by
\begin{equation}
	\begin{split}
		&\left(F^V_r(\theta^{(A)}_m, \phi^{(A)}_m)\quad F^H_r(\theta^{(A)}_m, \phi^{(A)}_m)\right) \\
		&\cdot\begin{pmatrix}
			\exp\left[\jmath \psi_m^{VV}\right] & \cfrac{1}{\sqrt{\kappa_m}}\exp\left[\jmath \psi_m^{HV}\right] \\
			\cfrac{1}{\sqrt{\kappa_m}}\exp\left[\jmath \psi_m^{VH}\right] & \exp\left[\jmath \psi_m^{HH}\right]
		\end{pmatrix}\\
		&\cdot\left(F^V_s(\theta^{(D)}_m, \phi^{(D)}_m)\quad F^H_s(\theta^{(D)}_m, \phi^{(D)}_m)\right)^T.
	\end{split}
\end{equation}
The matrix structure is now obtained as follows.
Let us define the polarized spatial transmit matrices $\mathbf S^V\in \mathbb C^{S \times M}$ and $\mathbf S^H\in \mathbb C^{S \times M}$ as
\begin{align}
	S^V_{s, m} &= F^V_s(\theta^{(D)}_m, \phi^{(D)}_m) \exp [\jmath k_0 \mathbf p_s^T \mathbf d_m] \\
	S^H_{s, m} &= F^H_s(\theta^{(D)}_m, \phi^{(D)}_m) \exp [\jmath k_0 \mathbf p_s^T \mathbf d_m].
\end{align}
Define the spatial receive matrices $\mathbf R^V\in \mathbb C^{R \times M}$ and $\mathbf R^H\in \mathbb C^{R \times M}$ similarly.
Using a straightforward definition for $\bm \psi^{VV}$, $\bm \psi^{VH}$, $\bm \psi^{HV}$ and $\bm \psi^{HH}$, as well as the depolarization vector $\bm \kappa = \left[1/\sqrt{\kappa_1}, \ldots, 1/\sqrt{\kappa_M}\right]^T$ we can define the polarized channel matrices $\mathbf H^{VV}$, $\mathbf H^{HV}$, $\mathbf H^{VH}$ and $\mathbf H^{HH}$. Each may be computed as e.g.
\begin{align*}
	\mathbf H^{HV}(f,t) &= \mathbf R^H \mathbf{diag} \, \left( \bm \kappa \odot \bm \psi^{HV} \odot \bm \rho \odot \bm \nu(t) \odot \bm \xi(f) \right) \left(\mathbf S^V\right)^T
\end{align*}
The final channel matrix can then be computed as
\begin{equation}
	\begin{split}
		\mathbf H(f,t) &= \mathbf H^{VV}(f,t) + \mathbf H^{VH}(f,t) \\
		&+ \mathbf H^{HV}(f,t) + \mathbf H^{HH}(f,t).
	\end{split}
\end{equation}
There are now 4 matrices to store for each link instead of 2---the transmit and receive array responses in both polarization domains.
The 4 polarized channel matrices can then be efficiently computed and summed to obtain the final channel for any time and frequency point.
\bibliographystyle{IEEEtran}
\bibliography{IEEEabrv,cm_awpl}

\end{document}